# On the Beamforming Design for Efficient Interference Alignment

Sang Won Choi, *Student Member, IEEE*, Syed A. Jafar, *Member, IEEE*, and Sae-Young Chung, *Senior Member, IEEE*

*Abstract*—An efficient interference alignment (IA) scheme is developed for $K$-user single-input single-output frequency selective fading interference channels. The main idea is to steer the transmit beamforming matrices such that at each receiver the subspace dimensions occupied by interference-free desired streams are asymptotically the same as those occupied by all interferences. Our proposed scheme achieves a higher multiplexing gain at any given number of channel realizations in comparison with the original IA scheme, which is known to achieve the optimal multiplexing gain asymptotically.

## I. INTRODUCTION

As an effective technique for interference management, interference alignment (IA)[1] has been proposed to achieve the optimal multiplexing gain asymptotically in single-input single-output (SISO) multi-user fading interference channel (IC) [2]. Subsequently, the IA scheme has been further studied in an explicit manner, which can be classified into two categories: One is to achieve the IA in signal scale [3]-[5]. Specifically, multi-user interferences at each receiver are aligned based on carefully constructed signal structures. The other is to align interferences in signal space [6], [7]. In this approach, transmit beamforming technique is used to align the multi-user interferences, which are nullified by zero-forcing (ZF) at each receiver.

The main contribution of this paper is to propose an efficient IA scheme in signal space when the number of users is greater than or equal to 3. Specifically, we propose a transmit beamforming design criterion so that a strictly higher multiplexing gain is attained in comparison with the original IA scheme [2] at any given number of channel realizations.

## II. SYSTEM MODEL

We consider the following channel model:

$$\mathbf{Y}^{[k]} = \sum_{l=1}^{K} \mathbf{H}^{[kl]} \mathbf{X}^{[l]} + \mathbf{Z}^{[k]}, \ \forall k \in \mathcal{K} = \{1, 2, \cdots, K\}, \quad (1)$$

where $M$ is the number of available frequency selective channel realizations, $\mathbf{Y}^{[k]} \in \mathbb{C}^M$ ($\mathbf{X}^{[l]} \in \mathbb{C}^M$) is the received (transmitted) signal vector of size $M \times 1$ at the $k$-th receiver (transmitter), and $\mathbf{H}^{[kl]} \in \mathbb{C}^{M \times M}$ is the $M \times M$ diagonal channel matrix from the $l$-th transmitter to the $k$-th receiver. Here, the $(i,i)$-th component $h^{[kl]}(i)$ of the channel matrix $\mathbf{H}^{[kl]}$ is the channel coefficient in the $i$-th frequency channel realization. We assume the channel is not time-varying. The noise vector $\mathbf{Z}^{[k]} \in \mathbb{C}^M$ is complex Gaussian with zero mean and the covariance of $\mathbf{I}_M$, where $\mathbf{I}_M$ is the identity matrix of size $M \times M$.

We assume the following in this paper.
- All transmitters and receivers are equipped with a single antenna.
- All $\mathbf{H}^{[kl]}$'s are known in advance at all the transmitters and all the receivers.
- The message of the $i$-th transmitter is independent from that of the $j$-th transmitter, $\forall i, j \in \mathcal{K}$ with $i \neq j$.
- All diagonal components of $\mathbf{H}^{[kl]}$'s are drawn independent and identically distributed (i.i.d.) from a continuous distribution, and absolute values of all diagonal elements are bounded below by a nonzero minimum value and above by a finite maximum value.

*Remark 1:* In SISO frequency selective fading IC, it is possible to use multiple frequency selective channel instances in a combined manner, which leads to diagonal channel matrices $\mathbf{H}^{[kl]}$'s, $\forall k, l \in \mathcal{K}$.

The multiplexing gain $r$ [8] of the $K$-user IC is defined as

$$r = \lim_{\text{SNR} \to \infty} \frac{R_+(\text{SNR})}{\log(\text{SNR})}, \quad (2)$$

where $R_+(\text{SNR})$ is an achievable sum-rate[2] at signal-to-noise ratio (SNR), where SNR is defined as the total power across all transmitters.

## III. PRELIMINARIES

The essence of the IA scheme lies in design of transmit beamforming matrices. Basically, interferences are aligned at each receiver using the designed beamforming matrices to maximize the total number of interference-free desired streams.

In the original IA scheme, the following IA conditions were proposed [2].

$$\mathbf{H}^{[1i]}\mathbf{V}^{[i]} = \mathbf{H}^{[13]}\mathbf{V}^{[3]}, \ \forall i \in \mathcal{K} \backslash \{1,3\} \quad (3)$$

and

$$\mathbf{H}^{[jk]}\mathbf{V}^{[k]} \prec \mathbf{H}^{[j1]}\mathbf{V}^{[1]}, \forall j \in \mathcal{K} \backslash \{1,k\}, \ \forall k \in \mathcal{K} \backslash \{1\}, \quad (4)$$

where $\mathbf{V}^{[k]}$ is a $M \times d^{[k]}$ transmit beamforming matrix of the $k$-th user. Here, $d^{[k]}$ is the number of streams of the $k$-th user. Note that in (1) we have

$$\mathbf{X}^{[k]} = \mathbf{V}^{[k]}\mathbf{S}^{[k]}, \quad (5)$$

---

[1]The first implicit IA scheme has been studied in [1].

[2]The definitions of achievable rate and capacity region follow from [2], which is omitted due to space limitations.

TABLE I
INTERFERENCE ALIGNMENT CONDITIONS

| The first Rx. $(i \in \mathcal{K}\backslash\{1,3\})$ | The second Rx. | The $k$-th Rx. $k \in \mathcal{K}\backslash\{1,2\}$ |
|---|---|---|
| $\mathbf{V}^{[i]} = \left(\mathbf{H}^{[1i]}\right)^{-1} \mathbf{H}^{[13]}$ $\cdot \mathbf{V}^{[3]}$. | $\mathbf{T}_3^{[2]} \mathbf{V}^{[3]} \prec \mathbf{V}^{[1]}$, $\mathbf{T}_4^{[2]} \mathbf{V}^{[3]} \prec \mathbf{V}^{[1]}$, $\vdots$ $\mathbf{T}_K^{[2]} \mathbf{V}^{[3]} \prec \mathbf{V}^{[1]}$. | $\mathbf{T}_2^{[k]} \mathbf{V}^{[3]} \prec \mathbf{V}^{[1]}$, $\mathbf{T}_3^{[k]} \mathbf{V}^{[3]} \prec \mathbf{V}^{[1]}$, $\vdots$ $\mathbf{T}_{k-1}^{[k]} \mathbf{V}^{[3]} \prec \mathbf{V}^{[1]}$, $\mathbf{T}_{k+1}^{[k]} \mathbf{V}^{[3]} \prec \mathbf{V}^{[1]}$, $\mathbf{T}_{k+2}^{[k]} \mathbf{V}^{[3]} \prec \mathbf{V}^{[1]}$, $\vdots$ $\mathbf{T}_K^{[k]} \mathbf{V}^{[3]} \prec \mathbf{V}^{[1]}$. |

where $\mathbf{S}^{[k]}$ is a $d^{[k]} \times 1$ vector of streams, and $\mathbf{A} \prec \mathbf{B}$ means that the column space of $\mathbf{A}$ is included in that of $\mathbf{B}$.

Note that (3) means that all the interferences are aligned exactly to occupy the same subspace at receiver 1, while (4) implies that the subspace spanned by the first transmitter's interferences includes all the subspaces spanned by all the other transmitters' interferences.

We define $\mathbf{T}_j^{[k]}$'s as

$$\mathbf{T}_j^{[k]} = \left(\mathbf{H}^{[k1]}\right)^{-1} \mathbf{H}^{[kj]} \left(\mathbf{H}^{[1j]}\right)^{-1} \mathbf{H}^{[13]}, \ \forall j, \ k \in \mathcal{K}\backslash\{1\} \tag{6}$$

with $j \neq k$. Then, the IA conditions (3) and (4) are equivalently rewritten as in TABLE I. Note that with probability 1 all $\mathbf{T}_j^{[k]}$'s are full-rank and $\mathbf{T}_a^{[b]} \neq \mathbf{T}_c^{[d]}$ for $a \neq c$ or $b \neq d$ since all the channel coefficients are assumed to be drawn i.i.d. from a continuous distribution.

Suppose that there exist $\mathbf{V}^{[k]}$'s ($\forall k \in \mathcal{K}$) satisfying the IA conditions in TABLE I, where the number of column vectors in $\mathbf{V}^{[k]}$ is $d^{[k]}$. When each transmitter transmits streams using the corresponding transmit beamforming matrix $\mathbf{V}^{[k]}$ and each receiver decodes the desired streams by ZF, the multiplexing gain $r$ becomes

$$r = \frac{(K-1)d^{[3]} + d^{[1]}}{d^{[3]} + d^{[1]}}, \tag{7}$$

where $d^{[2]} = d^{[3]} = \cdots = d^{[K-1]} < d^{[1]}$ from (3) and (4).

*Remark 2:* When $d^{[1]}/d^{[3]}$ approaches 1, (7) becomes close to the upper bound of $K/2$ on the optimal multiplexing gain [2], [9].

*Remark 3:* Note that at the $k$-th receiver ($\forall k \in \mathcal{K}$) the ZF is possible due to the fact that effective channel matrix of size $(d^{[3]} + d^{[1]}) \times (d^{[3]} + d^{[1]})$ is full-rank with probability 1 [2], where the effective channel matrix is formed by the channel matrices $\mathbf{H}^{[kl]}$'s and beamforming matrices $\mathbf{V}^{[l]}$'s, $\forall l = 1, 2, \cdots, K$.

In this paper, we propose an efficient beamforming design satisfying the IA conditions in TABLE I.

## IV. PROPOSED INTERFERENCE ALIGNMENT SCHEME

We propose the following beamforming design criterion for our IA scheme. Assume a nonnegative integer $n^*$ is given,

then we define the following:

$$\mathbf{V}^{[3]} = \left\{ \left(\mathbf{T}_3^{[2]}\right)^{-1} \prod_{k,\ l \in \mathcal{K}\backslash\{1\},\ k \neq l,\ (k,l) \neq (2,3)} \left(\left(\mathbf{T}_3^{[2]}\right)^{-1} \mathbf{T}_l^{[k]}\right)^{n_{kl}} \right.$$

$$\left. \cdot \mathbf{1}_M \left| \sum_{k,\ l \in \mathcal{K}\backslash\{1\},\ k \neq l,\ (k,l) \neq (2,3)} n_{kl} \leq n^* \right. \right\} \tag{8}$$

and

$$\mathbf{V}^{[1]} = \left\{ \prod_{k,\ l \in \mathcal{K}\backslash\{1\},\ k \neq l,\ (k,l) \neq (2,3)} \left(\left(\mathbf{T}_3^{[2]}\right)^{-1} \mathbf{T}_l^{[k]}\right)^{n_{kl}} \right.$$

$$\left. \cdot \mathbf{1}_M \left| \sum_{k,\ l \in \mathcal{K}\backslash\{1\},\ k \neq l,\ (k,l) \neq (2,3)} n_{kl} \leq n^* + 1 \right. \right\}, \tag{9}$$

where $n_{kl}$'s are nonnegative integers, and $\mathbf{1}_M$ is the all one vector of size $M \times 1$. The beamforming design criterion in (8) and (9) comes from the following motivation: It is reasonable to construct $\mathbf{V}^{[1]}$ and $\mathbf{V}^{[3]}$ such that $d^{[1]}/d^{[3]}$ becomes close to 1, which is confirmed from (7). For this, we use the following strategy. First, each column of $\mathbf{V}^{[1]}$ and $\mathbf{V}^{[3]}$ has the form of multiplication of matrices $\left(\mathbf{T}_3^{[2]}\right)^{-1} \mathbf{T}_l^{[k]}$'s with exponents $n_{kl}$'s. Second, We limit the sum of $n_{kl}$'s to satisfy the IA conditions in TABLE I.

Note that column space of the beamforming matrix $\mathbf{V}^{[1]}$ in (9) always includes all column spaces of $\mathbf{T}_l^{[k]} \mathbf{V}^{[3]}$'s, $\forall k,\ l \in \mathcal{K}\backslash\{1\}$ with $k \neq l$ and other beamforming matrices $\mathbf{V}^{[k]}$'s ($\forall k \in \mathcal{K}\backslash\{1, 3\}$) are constructed directly from $\mathbf{V}^{[3]}$ using (3). Thus, the IA conditions in TABLE I are satisfied by following (8) and (9).

*Theorem 1:* When $K \geq 3$, the multiplexing gain of

$$\frac{(K-1)(n^*+1) + n^* + N + 1}{2n^* + N + 2} \tag{10}$$

is achievable for any nonnegative integer $n^*$ in the $K$-user SISO fading IC, where $N = (K-1)(K-2) - 1$.

*Proof:* It is sufficient to specify $d^{[3]}$ and $d^{[1]}$ to get the achievable multiplexing gain from (7). Suppose that we follow the beamforming design criterion in (8) and (9), which satisfies the IA conditions in TABLE I. Since the construction rule for the beamforming matrix $\mathbf{V}^{[3]}$ is exactly the same as that for $\mathbf{V}^{[1]}$, it is enough to show the number $d^{[3]}$ of column vectors in $\mathbf{V}^{[3]}$ satisfying (8). From basic combinatoric results [11], $d^{[3]}$ is easily derived to be

$$d^{[3]} = \sum_{n=0}^{n^*} \binom{n+N-1}{N-1}$$

$$= \binom{n^*+N}{N}. \tag{11}$$

Similarly, we get

$$d^{[1]} = \binom{n^*+N+1}{N}. \tag{12}$$

Finally, we obtain

$$\begin{aligned} r &= \frac{(K-1) + d^{[1]}/d^{[3]}}{1 + d^{[1]}/d^{[3]}} \\ &= \frac{(K-1)(n^*+1) + n^* + N + 1}{2n^* + N + 2}, \end{aligned} \qquad (13)$$

which completes the proof. ∎

*Remark 4:* In the sense of the high-SNR offset, the proposed IA scheme can be improved by using a better vector instead of $\mathbf{1}_M$ as in [10] while maintaining the same achievable multiplexing gain in *Theorem* 1.

*Remark 5:* Note that the proposed $\mathbf{V}^{[3]}$ and $\mathbf{V}^{[1]}$ lead to full-rank effective channel matrices at each receiver (with probability 1), which can be shown using contradiction in a similar manner as in [2]. This is because all the channel coefficients are drawn i.i.d. from a continuous distribution.

*Remark 6:* As $n^*$ tends to infinity, the optimal multiplexing gain of $K/2$ is achieved asymptotically since $d^{[1]}/d^{[3]}$ tends to 1 from (11) and (12).

*Remark 7:* In [2], the ratio $d^{[1]}/d^{[3]}$ has been shown to be $(m+2)^N/(m+1)^N$ for a nonnegative integer $m$, which corresponds to the case when $M = (m+1)^N + (m+2)^N$ in (1). Given the fixed number of channel realizations, i.e., $M = (m+1)^N + (m+2)^N = 2n^* + N + 2$, the ratio $d^{[1]}/d^{[3]}$ coming from the proposed beamforming design satisfies

$$\begin{aligned} \frac{d^{[1]}}{d^{[3]}} &= \frac{n^* + N + 1}{n^* + 1} \\ &= 1 + \frac{2N}{(m+1)^N + (m+2)^N - N} \\ &\leq \frac{(m+2)^N}{(m+1)^N}, \end{aligned} \qquad (14)$$

where the equality holds iff $N = 1$, which is equivalent to $K = 3$.

*Remark 8:* The proposed beamforming design criterion in (8) and (9) is more efficient than the original one [2] when $K \geq 4$ since beamforming matrices are designed such that $d^{[1]}/d^{[3]}$ becomes closer to 1 while satisfying the IA conditions (3) and (4), which is shown from (14).

*Corollary 1:* When $K \geq 3$, the multiplexing gain of

$$\frac{(KM-1)(n^*+1) + n^* + N' + 1}{2n^* + N' + 2} \qquad (15)$$

is achievable in a fading IC with $K$ transmitters and $K$ receivers equipped with $M$ antennas each. Here, $N' = (KM-1)(KM-2) - 1$.

*Proof:* It follows directly from *Theorem* 1 by interpreting $K$-user fading IC with $M$ antennas as $KM$-user fading IC with a single antenna. ∎

## V. NUMERICAL RESULTS

In Fig. 1, the achievable multiplexing gains are shown for the proposed IA scheme and for the original one when $K = 4$. It is observed that for each number of channel uses, the proposed one attains a strictly higher multiplexing gain than the original one. Note that this becomes more significant as $K$ increases beyond 3.

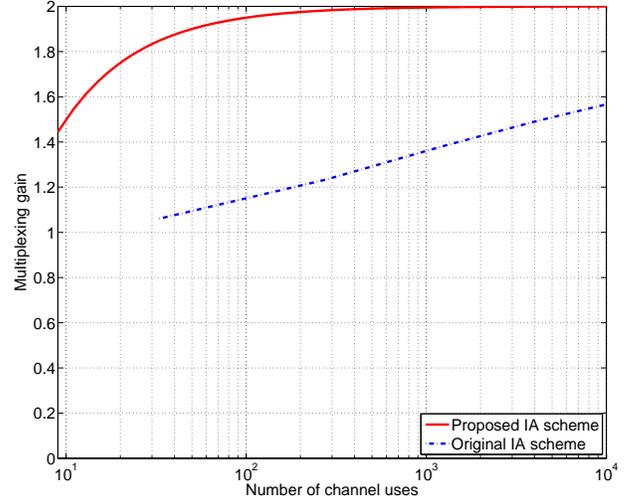

Fig. 1. Achievable multiplexing gains of the proposed IA scheme and the original IA scheme [2] when $K = 4$.

## REFERENCES

[1] M. Maddah-Ali, A. Motahari, and A. Khandani, "Signaling over MIMO multi-base systems - combination of multi-access and broadcast schemes," in *Proc. IEEE Int. Symp. Inform. Theory*, 2006.
[2] V. R. Cadambe and S. A. Jafar, "Interference alignment and the degrees of freedom for the K user interference channel," *IEEE Trans. Inf. Theory,* vol. IT-54, no. 8, pp. 3425-3441, Aug. 2008.
[3] V. R. Cadambe, S. A. Jafar, and S. Shamai (Shitz), "Interference alignment on the deterministic channel and application to fully connected AWGN interference networks," *IEEE Trans. Inf. Theory,* vol. IT-55, no. 1, pp. 269-274, Jan. 2009.
[4] S. Sridharan, A. Jafarian, S. Vishwanath, S. A. Jafar, and S. Shamai (Shitz), "A layered lattice coding scheme for a class of three user Gaussian interference channels," preprint.
[5] R. Etkin and E. Ordentlich, "On the degrees-of-freedom of the $K$-user Gaussian interference channel," preprint.
[6] K. Gomadam, V. R. Cadambe, and S. A. Jafar, "Approaching the capacity of wireless networks through distributed interference alignment," preprint.
[7] T. Gou and S. A. Jafar, "Degrees of freedom of the $K$ user $M \times N$ MIMO interference channel," preprint.
[8] L. Zheng and D. N. C. Tse, "Diversity and multiplexing: A fundamental tradeoff in multiple-antenna channels," *IEEE Trans. Inf. Theory,* vol. 49. no. 5, pp. 1073-1096, May 2003.
[9] A. Høst-Madsen and A. Nosratinia, "The multiplexing gain of wireless networks," in *Proc. IEEE Int. Symp. Inform. Theory*, 2005.
[10] M. Shen, A. Høst-Madsen, and J. Vidal, "An improved interference alignment scheme for frequency selective channels," in *Proc. IEEE Int. Symp. Inform. Theory*, 2008.
[11] J. H. van Lint and R. M. Wilson, *A course in combinatorics,* Cambridge univ. press, 1992.